\documentclass{article}
\usepackage{amsmath}
\usepackage[final]{neurips_2026}
\usepackage[utf8]{inputenc} 
\usepackage[T1]{fontenc}    
\usepackage{hyperref}       
\usepackage{url}            
\usepackage{booktabs}       
\usepackage{amsfonts}       
\usepackage{nicefrac}       
\usepackage{microtype}      
\usepackage{xcolor}         
\usepackage{graphicx}
\usepackage{tikz}
\usetikzlibrary{positioning, arrows.meta, fit, calc}
\usepackage{float}
\usepackage{hyperref}
\usepackage{url}
\usepackage{booktabs}
\usepackage{amsfonts}
\usepackage{nicefrac}
\usepackage{microtype}
\usepackage{listings}
\usepackage{multirow}
\usepackage{capt-of}
\usepackage{tcolorbox}
\usepackage{xcolor}
\usepackage{enumitem}
\usepackage{siunitx}
\usepackage{rotating} 
\usepackage{booktabs}
\usepackage[table]{xcolor}
\usepackage{placeins}
\usepackage{caption}

\tcbuselibrary{skins,breakable}
\usepackage{tcolorbox}
\newtcolorbox{qualbox}[1]{
  colback=gray!3, colframe=gray!40, boxrule=0.5pt, arc=2pt,
  left=6pt, right=6pt, top=4pt, bottom=4pt,
  fonttitle=\bfseries\small, title=#1
}

\usepackage{xcolor}
\usepackage{booktabs}
\definecolor{deltagreen}{HTML}{2E7D32}
\definecolor{deltared}{HTML}{C62828}
\newcommand{\up}[1]{{\scriptsize\color{deltagreen}$+#1$}}

\lstdefinestyle{promptstyle}{
    basicstyle=\ttfamily\footnotesize,
    breaklines=true,
    breakatwhitespace=true,
    breakindent=0pt,
    columns=fullflexible,
    frame=single,
    framesep=6pt,
    rulecolor=\color{gray!50},
    backgroundcolor=\color{gray!5},
    showstringspaces=false,
    keepspaces=true,
    upquote=true,
    literate={`}{{\textasciigrave}}1,
    aboveskip=10pt,
    belowskip=10pt,
}

\title{FinRegQA-EU: Corruption-Based Preference Data for Grounded EU Financial Regulatory Question Answering}

\author{%
  Aulia Kharis Rakhmasari\thanks{Main contributor. Correspondence to \texttt{arakhmasari@ethz.ch}.} , 
   Fan Yu, Alexander Hoyle, Elliot Ash \\
  ETH Z\"urich \\
  Z\"urich, Switzerland \\
}

\begin{document}

\maketitle


\begin{abstract}

Large Language Models (LLMs) struggle with region-specific factual knowledge, particularly in financial regulation. While benchmarks such as CFinBench \citep{ding_cnfinbench_2025} provide broad coverage of financial knowledge in other regions, no comparable resource exists for European financial regulation. We close this gap by proposing an end-to-end pipeline for evaluating and improving LLMs on European financial regulatory question answering. Our dataset is grounded in the official Q\&A corpora of the European Banking Authority (EBA) and the European Securities and Markets Authority (ESMA).


We evaluate candidate answers with a pointwise LLM-as-a-Judge protocol using three judges from distinct model families, retain only unanimously judged pairs, and sharpen the rejected side through a taxonomy of regulatory failure modes : law swaps, article swaps, hallucinated citations, and hallucinated text. By fine-tuning on the resulting preference pairs exposes a divergence at the core of our findings: supervised fine-tuning attains the highest judge score of any method yet got the lowest rule-based Citation F1. SFT is producing longer answers with three times as many citations, most of them unsupported. DPO and GRPO offer the better trade-off with more concise answers and the highest citation F1 among fine-tuned models. Standard LLM-as-a-Judge evaluation rewards citation density as evidence of grounding and cannot detect when those citations are fabricated. Thus, we detect a blind spot that matters wherever answers must be verifiable. We release the benchmark and evaluation stack at \url{https://github.com/auliakharis/FinRegQA-EU}.

\end{abstract}
\section{Introduction}

Large Language Models (LLMs) are increasingly deployed across the financial sector, including in regulator-facing workflows where outputs must align with the governing regulatory text. Existing evaluations, however, focus largely on textbook-style question answering and numerical problem solving. \citet{chen_finqa_2022} target deep questions over financial documents requiring complex numerical reasoning, and \citet{chen_convfinqa_2022} extend this to conversational, multi-turn settings. Related benchmarks \citep{islam_financebench_2023, wang_fingpt_2023} likewise centre on extraction and arithmetic skills. Legal benchmarks \citep{cao_safelawbench_2025, guha_legalbench_2023, li_legalagentbench_2025} address the legal domain at a generic level, but do not isolate financial regulation. 

Most recently, CNFinBench \citep{ding_cnfinbench_2025} closes part of this gap by evaluating LLMs along capability, compliance, and safety axes, and by introducing multi-turn adversarial tasks. Its results show that most systems achieve only partial resistance under adversarial testing, indicating that current models remain unreliable without cited, verifiable reasoning. CNFinBench, however, is built entirely around the Chinese regulatory regime. No comparable resource exists for Europe, despite the fact that the EU publishes an extensive body of open regulatory Q\&A material through the European Banking Authority (EBA) and the European Securities and Markets Authority (ESMA).

We close this gap by proposing an end-to-end pipeline for evaluating and improving LLMs on EU financial regulatory question answering, grounded in the official Q\&A corpora of the EBA and ESMA. To measure how well current models answer regulatory questions, we use a pointwise LLM-as-a-Judge protocol that scores answers along four dimensions: \emph{accuracy} against the official answer, \emph{completeness} of key points, \emph{topic coherence} with the stated subject matter, and \emph{citation quality} in terms of the specificity and correctness of legal references. We employ multiple judge models, evaluate each independently, and use the resulting scores to construct preferred and less-preferred answer pairs. To sharpen the separation between the two, we synthesise less-preferred answers through a taxonomy of regulatory failure modes including article swap, law swap, and hallucination injection. Because our aim is to improve models on their own output distribution, we fine-tune on these pairs and evaluate against the base model.

\paragraph{Contributions.}
\begin{itemize}
    \item \textbf{An EU regulatory QA evaluation benchmark} built from
    authoritative EBA and ESMA Q\&A material, scored along four
    interpretable dimensions: accuracy, completeness, topic coherence, and
    citation quality.
    \item \textbf{A regulatory failure-mode taxonomy} including article swap, law
    swap, and hallucination injection for synthesising calibrated
    less-preferred answers, yielding preference-learning datasets that
    target the specific errors LLMs make in this regime.
\item \textbf{A controlled comparison} of four fine-tuning methods (SFT, DPO, Dr.~DPO, GRPO) trained on the full preference set, together with a per-corruption DPO ablation isolating the contribution of each failure mode. All models are evaluated on both judge scores and rule-based citation F1 against the official answers.
\end{itemize}
\section{Related Work}

\paragraph{Legal and financial QA benchmarks.}
Most financial benchmarks test numerical reasoning: reading tables,
extracting numbers, doing arithmetic \citep{chen_finqa_2022,
chen_convfinqa_2022, islam_financebench_2023, wang_fingpt_2023}. Legal
benchmarks \citep{guha_legalbench_2023, li_legalagentbench_2025,
cao_safelawbench_2025} cover law in general, but financial regulation is
only a small part of what they test. The closest work to ours is
CNFinBench \citep{ding_cnfinbench_2025}, which does focus on financial
regulatory compliance but only for China. Because regulation is jurisdiction-specific by constructions and currently, no equivalent resource exists for European Union. We close this gap by introducing a benchmark for EU Financial Regulatory QA grounded in the official Q\&A material published by the EBA and ESMA.

\paragraph{LLM-as-a-Judge.}
Using one LLM to score another model's answers is now standard practice, and prior work reports that judges agree with humans about as well as humans agree with each other \citep{zheng_judging_2023, licht_measuring_2025, wang_improving_2025}. That work also documents the method's failure in 
position bias (favoring whichever answer appears first), verbosity bias, and self-preference \citep{zheng_judging_2023, zeng_evaluating_2024}.
These biases affect pairwise comparison in particular, which is one reason we score answers one at a time rather than in pairs. Pointwise scoring also gives us something pairwise does not which is a scalar score per dimension that can be the threshold for building preference pairs.

\paragraph{Judge agreement.} We quantify agreement across the judge ensemble
with Krippendorff's $\alpha$~\citep{krippendorff_content_2004}, which measures
observed disagreement against the disagreement expected by chance:
\begin{equation}
\alpha = 1 - \frac{D_o}{D_e},
\end{equation}
where $D_o$ is the observed disagreement among judge verdicts and $D_e$ the
disagreement expected under random assignment. Values of $\alpha = 1$ indicate
perfect agreement and $\alpha = 0$ chance-level agreement. Unlike simple
percent agreement, $\alpha$ corrects for chance, handles more than two raters,
and accommodates missing verdicts, making it appropriate for our
three-judge ensemble.

\paragraph{Citation grounding and attribution.}
A common LLMs failure is citations hallucination. A model can produce a fluent answer and attach a citation that does not
support it. This problem has been studied under some past research, \citet{liu_just_2021} formalize what it means for a
statement to be supported by a source, and \citet{gao_enabling_2023} build a
benchmark for text generation with citations. The finding most relevant to
us is that retrieval does not solve the problem.
\citet{liu_evaluating_2023} also show that commercial generative search engines which retrieve documents before answering still frequently produce
citations that do not back up their claims.
\citet{onweller_cited_2026} report the same pattern for frontier models,
with only 39--77\% of cited claims verified correct, and find that adding
retrieval does not close the gap. Similar issues appear in academic
citation verification \citep{sadeghi_deepsciverify_2026}.

Thus, it is clear that if more retrieval does not produce better citations, then by fixing the model, we can get better output compared to fixing the pipeline. That is why we fine-tune the model to produce a better citation-grounded output.

\paragraph{Preference optimisation.} Direct Preference Optimization (DPO) trains a model directly on pairs of preferred and rejected responses, without a separate reward model \citep{rafailov_direct_2024}. This method has been used mainly to align open-weight models. DPO learns little when rejected answer is never produced by the model, since the gradient pushes probability away. 

\paragraph{Bradley-Terry Model} \citep{bradley_rank_1952} is a statistical model for the pairwise comparions. Each item $i$ has a latent scalar $s_i$ and the probability that $i$ beats $j$ depends only on : 
\begin{equation}
P(i \succ j) \;=\; \frac{\exp(s_{i})}{\exp(s_{i}) + \exp(s_{j})} \;=\; \sigma(s_{i} - s_{j})
\end{equation}

The responses divided into preferred responses $y_{w}$ and rejected responses $y_{l}$. The latent scalar is the reward $r^*(x, y)$. Thus,  $P^*(y_{w} \succ y_{l} \mid x) = \sigma\big(r^*(x, y_{w}) - r^*(x, y_{l})\big)$. 

Maximizing the log-likelihood of the human-preferences under this model gives the standard reward-model loss : 
\begin{equation}
\mathcal{L}_{R} \;=\; -\,\mathbb{E}\big[ \log \sigma\big( r_{\phi}(x, y_{w}) - r_{\phi}(x, y_{l}) \big) \big]
\end{equation}

where $\sigma(\cdot)$ is the sigmoid functions. As the size of datasets $\mathcal{O}$ grows, the empirical distribution of dataset $\mathcal{O}$ converges to $p^*$ and the reward model $r_{\phi}$ will converge to the true reward model $r^*$.

\paragraph{Reinforcement Learning from Human Feedback (RLHF)}\citep{ouyang_training_2022} The standard RLHF procedure is divided into three phases, i) Supervised Fine Tuning (SFT), ii) reward modeling, and iii) Reinforcement Learning fine tuning. The main objective of RLHF paradigm is to optimize this following objecting using the reward model learned from the reward modeling. 

\begin{equation}
\max_{\pi_{\theta}} \; \mathbb{E}_{x \sim \mathcal{D},\, y \sim \pi_{\theta}(y \mid x)} \big[ r_{\phi}(x, y) \big] - \beta \, \mathbb{D}_{\mathrm{KL}} \big[ \pi_{\theta}(y \mid x) \,\|\, \pi_{\mathrm{ref}}(y \mid x) \big] \label{eq:rlhf}
\end{equation}

\paragraph{DPO} \citep{rafailov_direct_2024} DPO offers an alternative of the RLHF pipeline (train a reward model then optimize it with PPO) with a single supervised loss on preference pairs. 

From the \eqref{eq:rlhf}, the optimal policy under a KL-constrained reward objective is 
\begin{equation}
\pi^{*}(y \mid x) \;\propto\; \pi_{\mathrm{ref}}(y \mid x) \exp\big(r(x, y)/\beta\big) \label{eq:dpo_reward}
\end{equation}
By inverting \eqref{eq:dpo_reward}, the reward is expressible in terms of the policy itself:
\begin{equation}
r(x, y) \;=\; \beta \log \frac{\pi_{\theta}(y \mid x)}{\pi_{\mathrm{ref}}(y \mid x)} \;+\; \beta \log Z(x) \label{eq:dpo_reward2}
\end{equation}
By plugging \eqref{eq:dpo_reward2} into the Bradley--Terry preference likelihood and the partition
function $Z(x)$ cancels, it is identical for $y_{w}$ and $y_{l}$ given the
same prompt. Thus, the DPO loss is given by 
\begin{equation}
\mathcal{L}_{\mathrm{DPO}} \;=\; -\,\mathbb{E}\left[
\log \sigma\!\left(
\beta \log \frac{\pi_{\theta}(y_{w} \mid x)}{\pi_{\mathrm{ref}}(y_{w} \mid x)}
\;-\;
\beta \log \frac{\pi_{\theta}(y_{l} \mid x)}{\pi_{\mathrm{ref}}(y_{l} \mid x)}
\right)\right]
\end{equation}

\paragraph{Distributionally Robust Optimization (DRO)} \citep{duchi_learning_2020}. DRO provides a framework to mitigate the uncertainty in the training data. It achieves this by optimizing the worst-case expected loss across all set of potential distributions $Q$. While the standard ERM only minimizes expected loss, DRO replaces it with minimizes the worst loss over a set of distributions which are considered plausible. 

\textbf{Standard ERM:}
\begin{equation}
\min_{\theta} \; \mathbb{E}_{(x,y) \sim P_{\mathrm{train}}}\big[ \ell(\theta; x, y) \big]
\end{equation}

\textbf{DRO:}
\begin{equation}
\min_{\theta} \; \sup_{Q \in \mathcal{U}(P_{\mathrm{train}})} \; \mathbb{E}_{(x,y) \sim Q}\big[ \ell(\theta; x, y) \big]
\end{equation}

Intuitively, DRO exhibit increased robustness due to the presence of $Q$ that acts as an adversary, optimizing the model under adversarial pertubations distribution. 

\paragraph{Distributionally Robustifying DPO (Dr.DPO)} \citep{wu_towards_2025} Building upon the principles of DRO, Dr.DPO designed to enchance DPO's resilience to pairwise noise while preserving its inherent robustness to pointwise noise. The objective is formulated as follows : 

\begin{equation}
\max_{\mathcal{O}'} \mathbb{E}_{(x, y_{w}, y_{l}) \sim \mathcal{O}'}\big[ h(x, y_{w}, y_{l}) \big]
\quad \text{s.t.} \quad \mathbb{D}_{\phi}(\mathcal{O}', \mathcal{O}) \leq \eta'
\end{equation}

$\mathbb{D}_{\phi}(\mathcal{O}', \mathcal{O}) $ denote the discrepancy between hypothethical distribution $\mathcal{O}'$ and the dataset distribution $\mathcal{O}$. While $\eta'$ signifies as robustness radius. Thus, the ultimate loss functions formulated as follows : 

\begin{equation}
\mathcal{L}_{\mathit{Dr.\,DPO}}(\pi_{\theta}; \pi_{\mathit{ref}})
= -\beta' \log \mathbb{E}_{\mathcal{O}}\!\left[ \exp\!\left( \frac{h_{\mathit{DPO}}(x, y_{w}, y_{l})}{\beta'} \right) \right]
\end{equation}

where $h_{\mathit{DPO}}$ represents the log-likelihood of the DPO framework, defined as : 

\begin{equation}
h_{\mathit{DPO}}(x, y_{w}, y_{l})
= \log \sigma\!\left( \beta \log \frac{\pi_{\theta}(y_{w} \mid x)}{\pi_{\mathit{ref}}(y_{w} \mid x)}
- \beta \log \frac{\pi_{\theta}(y_{l} \mid x)}{\pi_{\mathit{ref}}(y_{l} \mid x)} \right)
\end{equation}

with $\beta$ and $\beta'$ being regularization coefficient respectively. 

\paragraph{Group DRO} \citep{sagawa_distributionally_2020} This method is the modification of DRO. Group DRO significantly improves the worst-group accuracy at small cost in avarage accuracy. Group DRO prevent models from learning pre-specified spurious correlations.   

Group DRO takes the uncertainty set to be all mixtures over the groups,
$\mathcal{U} = \big\{ \sum_{g} q_{g} P_{g} : q \in \Delta_{G} \big\}$, giving
\begin{equation}
\min_{\theta} \; \max_{q \in \Delta_{G}} \; \sum_{g=1}^{G} q_{g} \, \mathcal{L}_{g}(\theta)
\;=\;
\min_{\theta} \; \max_{g \in \{1, \dots, G\}} \; \mathcal{L}_{g}(\theta)
\end{equation}

Groups with higher current loss receive greater weight in the subsequent descent step. This step is done by alternating exponentiated gradient ascent on $q$ and gradient descent on $\theta$.

\begin{equation}
q_{g}^{(t+1)} \;\propto\; q_{g}^{(t)} \exp\big( \eta \, \hat{\mathcal{L}}_{g}(\theta^{(t)}) \big),
\qquad
\theta^{(t+1)} \;=\; \theta^{(t)} - \alpha \nabla_{\theta} \sum_{g} q_{g}^{(t+1)} \mathcal{L}_{g}(\theta^{(t)})
\end{equation}

\paragraph{Group Robust Preference Optimization (GRPO)} \citep{ramesh_group_2024}
GRPO aim to measure the alignment of the reward model on the
\emph{worst-case group loss}:
\begin{equation}
\max_{g \in \mathcal{G}} \; \mathcal{L}_{R}(r; \mathcal{D}_{g}).
\end{equation}

The \emph{group robust preference optimization} (GRPO) objective for a
specified policy $\pi$:
\begin{equation}
\mathcal{L}_{\mathrm{GR}}(\pi) := \max_{g \in \mathcal{G}} \mathcal{L}_{\mathrm{DPO}}(\pi, \mathcal{D}_{g})
= \max_{g \in \mathcal{G}} \left( -\, \mathbb{E}_{(x_{g}, y_{w}, y_{l}) \sim \mathcal{D}_{g}}
\Big[ \log \big( \sigma( \beta h_{\pi}(x_{g}, y_{w}, y_{l}) ) \big) \Big] \right).
\end{equation}
Leveraging the equivalent formulation of maximizing over discrete set, the GRPO
problem becomes
\begin{equation}
\min_{\pi} \mathcal{L}_{\mathrm{GR}}(\pi) = \min_{\pi} \; \max_{\alpha \in \Delta_{K}}
\sum_{g=1}^{K} \alpha_{g} \left( -\, \mathbb{E}_{(x_{g}, y_{w}, y_{l}) \sim \mathcal{D}_{g}}
\Big[ \log \big( \sigma( \beta h_{\pi}(x_{g}, y_{w}, y_{l}) ) \big) \Big] \right),
\end{equation}

Optimization alternates gradient descent on $\theta$ with minor ascent on $\alpha$. Groups that historically suffered with high loss with get more weight and scaled by group size $N_g$ so small groups are not disadvataged.

\section{Datasets}
\label{sec:datasets}

\paragraph{Sources} We construct our evaluation corpus from the two official Q\&A repositories
maintained by the European Banking Authority (EBA) \citep{european_banking_authority_single_nodate} and the European
Securities and Markets Authority (ESMA) \citep{european_securities_and_markets_authority_questions_nodate}. Both regulators operate public
single-rulebook Q\&A platforms in which supervised entities submit
interpretive questions on specific provisions of EU financial regulation,
and the regulator (or the relevant joint committee) issues an authoritative
written answer. These answers functions as legal cited supervisory for compliance practice. 

The EBA Q\&A covers regulatory on the practical application or implementation of the banking, payment services, AML/CFT and other legislation that falls within the EBA’s remit. This includes the associated delegated and implementing acts, RTS, ITS, guidelines and recommendations. While ESMA Q\&A covers securities market and aim to ensure a consistent application of the EU Single Rulebook for financial services. Together they provide a controlled, expert-curated source of regulatory question-answer pairs that is, to our knowledge, the most direct ground-truth signal available
for EU financial regulatory QA.

\paragraph{Schema and Preprocessing} Each raw entry on both portals provides a question, an answer, and
structured metadata indicating the legal act under interpretation, the
topic area, and the subject matter and background of
the question. We parse each entry into a uniform schema with six fields: legal act, topic, subject matter, background, question and ground truth. The legal act field is normalised to its official CELEX-style
identifier when available (e.g., Regulation~(EU)~2022/2554 for DORA). The
topic label is retained as published by the regulator, since it
reflects the supervisor's own taxonomy. 

\begin{table}[h]
\centering
\caption{Dataset composition. Each source question is answered by two
answerer models and scored independently by three judges. }
\label{tab:dataset}
\begin{tabular}{@{}p{0.35\textwidth}r@{}}
\toprule
\textbf{Statistic} & \textbf{Count} \\
\midrule
Source questions   & 2{,}138 \\
\quad from EBA     & 1{,}868 \\
\quad from ESMA    & 270 \\
\bottomrule
\end{tabular}
\end{table}

\paragraph{Preference-Pair Construction}

For preference learning we construct a corpus of preferred /
rejected answer pairs from the training split. Candidate answers
are generated by two open-weight 70B answerer models, there are Apertus-70B- Instruct and Llama-3.3-70B-Instruct, each conditioned on the structured context fields using the prompt in Appendix~\ref{app:prompt-answerer}.
Every candidate is scored by the three-judge ensemble along \emph{accuracy}, \emph{completeness},
\emph{topic coherence}, and \emph{citation quality}; the four scores are
combined into a weighted quality score \(s \in [1,5]\) with higher weight
on accuracy and citation quality, the two dimensions most predictive of
regulatory utility.

\begin{table}[H]
\centering
\caption{Judgments per (answerer, judge) pair. GLM-4.7-Flash returns
fewer judgments due to truncation in thinking mode.}
\label{tab:judgments}
\begin{tabular}{lccc}
\toprule
& \multicolumn{3}{c}{\textbf{Judge}} \\
\cmidrule(lr){2-4}
\textbf{Answerer} & Qwen3.5-27B & Gemma-4-31B-it & GLM-4.7-Flash \\
\midrule
Llama-3.3-70B-Instruct & 2{,}138 & 2{,}138 & 2{,}099 \\
Apertus-70B-Instruct   & 2{,}138 & 2{,}138 & 2{,}084 \\
\bottomrule
\end{tabular}
\end{table}

A candidate enters the \emph{preferred} set if its weighted score satisfies
\(s \geq \tau_{\mathrm{pref}}\). To produce a corresponding
\emph{less-preferred} answer for each preferred candidate, we apply one
of four targeted corruption operations, each instantiating a distinct
failure mode observed in baseline answerer outputs:

\begin{itemize}
    \item \textbf{Article Swap.} A cited article within the correct legal
    act is replaced with a plausible but incorrect article from the same
    act.
    \item \textbf{Law Swap.} The legal act itself is replaced with an
    adjacent but incorrect act (e.g., MiFID~II~$\leftrightarrow$~MiFIR).
    \item \textbf{Citation Hallucination.} A correct citation is replaced with a
    superseded or not-yet-applicable version of the provision (e.g., a
    pre-amendment CRR article number, or an act not yet in force at the
    question date).
    \item \textbf{Text Injection Hallucination.} A fabricated but
    plausible-sounding claim (a non-existent article, an invented numeric
    threshold, or a made-up implementing decision) is inserted into an
    otherwise correct answer.
\end{itemize}

To ensure that each pair is distinguishable yet non-trivial, we retain
a pair only when the score gap exceeds a separability threshold
\(s_{\mathrm{pref}} - s_{\mathrm{neg}} \geq \tau_{\mathrm{sep}}\),
discarding pairs in which the corruption was too subtle for the judge to
detect or in which the original preferred candidate was itself borderline.
\section{Methods}
\label{sec:methods}

\begin{figure}[h]
\centering
\resizebox{\textwidth}{!}{%
\begin{tikzpicture}[
    font=\scriptsize,
    every node/.style={align=center},
    prep/.style={
        draw, thick, rounded corners=2pt,
        fill=purple!8, draw=purple!60!black,
        inner sep=3pt, minimum width=2.4cm
    },
    eval/.style={
        draw, thick, rounded corners=2pt,
        fill=teal!8, draw=teal!60!black,
        inner sep=3pt, minimum width=2.4cm
    },
    side/.style={
        draw, thick, rounded corners=2pt,
        fill=gray!8, draw=gray!60!black,
        inner sep=2pt, font=\tiny
    },
    arr/.style={-{Stealth[length=1.8mm,width=1.8mm]}, thick, draw=black!65},
    bypass/.style={-{Stealth[length=1.8mm,width=1.8mm]}, thick, dashed, draw=black!40},
    phasehdr/.style={font=\footnotesize\bfseries\itshape},
    phcolor1/.style={text=purple!50!black},
    phcolor2/.style={text=teal!50!black}
]

\node[phasehdr, phcolor1] at (0, 0.6) {Phase 1: Preference Preparation};

\node[prep, minimum width=1.15cm] (eba)  at (-0.65, 0) {\textbf{EBA}};
\node[prep, minimum width=1.15cm] (esma) at (0.65, 0)  {\textbf{ESMA}};

\node[prep] (questions) at (0, -0.9) {
    \textbf{Q\&A pairs} ($n{=}2{,}134$)
};

\node[prep, minimum width=1.15cm, font=\tiny] (apertus) at (-0.65, -1.9) {\textbf{Apertus}\\70B $\to y_A$};
\node[prep, minimum width=1.15cm, font=\tiny] (llama)   at (0.65, -1.9)  {\textbf{Llama-3.3}\\70B $\to y_B$};

\node[prep] (judges) at (0, -3.1) {
    \textbf{Pointwise judges}\\
    \tiny Gemma-4 $\cdot$ Qwen3.5 $\cdot$ GLM-4.7\\
    \tiny 4 rubric with score (1--5)
};

\node[side] (agreement) at (-2.3, -3.1) {
    Judge $\alpha$
};

\node[prep] (filter) at (0, -4.5) {
    \textbf{Preference filter}\\
    \tiny All judges agree + Strong Separation\\
    \tiny $2{,}134\to359$
};

\node[phasehdr, phcolor1] at (4.5, 0.6) {Phase 2: Training};

\node[prep] (corruption) at (4.5, -2.0) {
    \textbf{Preference Pair Construction}\\
    \textbf{
    via Targeted Corruption}\\
    \tiny Article / law swap, fake\\
    \tiny cite / claim
};

\node[prep] (dpo) at (4.5, -3.8) {
    \textbf{DPO fine-tuning}\\
    \tiny $\pi_\theta =$ Llama-3.2-3B + LoRA\\
    \tiny $\pi_{\mathrm{ref}}$ frozen $\cdot$ $\beta = 0.1$
};

\node[phasehdr, phcolor2] at (9.0, 0.6) {Phase 3: Evaluation};

\node[eval] (gen) at (9.0, -2.0) {
    \textbf{Answer generation}\\
    \tiny Baseline vs.\ DPO\\
    \tiny on $n{=}100$ held-out  \\
    \tiny from test datasets 
};

\node[eval, minimum width=1.15cm, font=\tiny] (evjudge) at (8.35, -3.3) {\textbf{Judges}\\\tiny};
\node[eval, minimum width=1.15cm, font=\tiny] (evcite)  at (9.65, -3.3) {\textbf{Regex}\\\tiny Parser};

\node[eval, minimum width=1.15cm, font=\tiny] (mjudge) at (8.35, -4.5) {\textbf{Scores}\\\tiny 4 rubric};
\node[eval, minimum width=1.15cm, font=\tiny] (mcite)  at (9.65, -4.5) {\textbf{F1 Citation}\\\tiny Deterministic};

\draw[gray!35, dashed] (2.2, 0.8) -- (2.2, -5.2);
\draw[gray!35, dashed] (6.8, 0.8) -- (6.8, -5.2);

\draw[arr] (eba.south)  -- (questions.north -| eba.south);
\draw[arr] (esma.south) -- (questions.north -| esma.south);
\draw[arr] (questions.south -| apertus.north) -- (apertus.north);
\draw[arr] (questions.south -| llama.north)   -- (llama.north);
\draw[arr] (apertus.south) -- (judges.north -| apertus.south);
\draw[arr] (llama.south)   -- (judges.north -| llama.south);
\draw[arr]    (judges.south) -- (filter.north);
\draw[bypass] (judges.west)  -- (agreement.east);

\draw[arr] (filter.east) -- ++(0.3,0) |- (corruption.west)
    node[pos=0.25, right, font=\tiny] {359 train};

\draw[arr] (corruption.south) -- (dpo.north);

\draw[arr] (dpo.east) -- ++(0.3,0) |- (gen.west);

\draw[bypass] (filter.south) -- ++(0,-0.6) -| (7.7, -0.8) -| (gen.north)
    node[pos=0.22, below, font=\tiny] {100 held-out};

\draw[arr] (gen.south -| evjudge.north) -- (evjudge.north);
\draw[arr] (gen.south -| evcite.north)  -- (evcite.north);
\draw[arr] (evjudge.south) -- (mjudge.north);
\draw[arr] (evcite.south)  -- (mcite.north);

\end{tikzpicture}%
}
\caption{End-to-end FinRegQA-EU framework}
\label{fig:framework}
\end{figure}
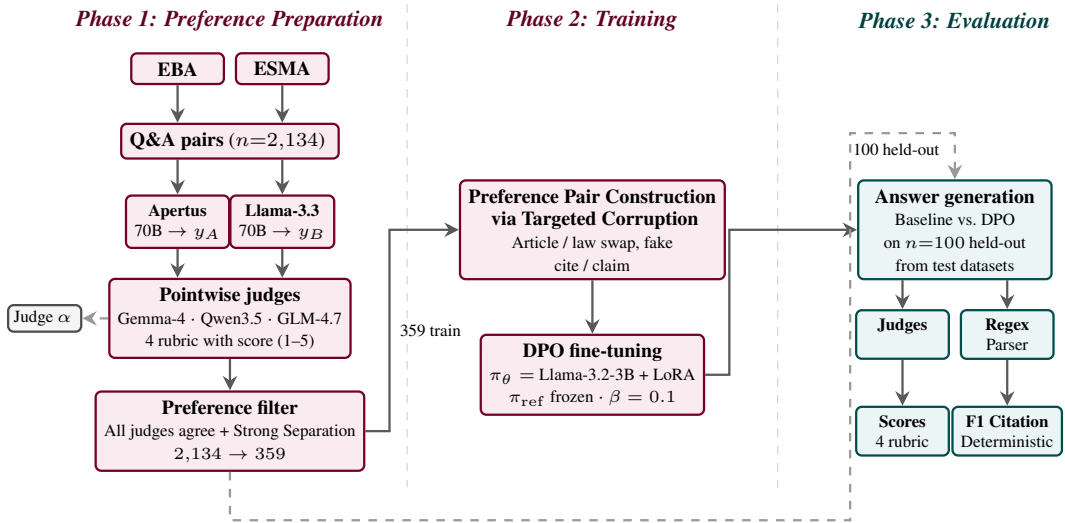

\paragraph{Phase~1} In the first phase, we construct preference pairs from the EBA and ESMA regulatory Q\&A corpora, resulting in a total of 2,134 question pairs. Each question has its own ground truth. To evaluate how well the models perform, we employ an LLM-as-a-judge approach on two large 70B models to generate answers $y_A$ and $y_B$. A pointwise ensemble of three judges scores each candidate answer on four dimensions (accuracy, completeness, coherence, and citation quality). From the two answers generated by the large models, we calculate which answer receives better evaluations based on the pointwise judges. We use three judge models here (Gemma-4, Qwen 3.5, and GLM 4.7), and we average their respective results.

With the results from these judges, we can derive which answer is preferred and which answer is less preferred based on the average score from each judge. These will later serve as the preference pairs. To ensure that the preference signal is strong between those two large models, we only select pairs where all three judges agree and show a strong separation (a difference of at least 1.5 in their scores).

\paragraph{Phase~2} Starting from the filtered set of strong preference pairs, we sharpen the contrast further by corrupting the less-preferred response in each pair. We apply five corruption strategies : article swap, law swap, hallucinated citations, and hallucinated text injections. On the resuting pairs, we fine-tune with four methods : SFT, DPO, Distributionally Robust DPO (Dr. DPO), and Group Robust Performance Optimization (GRPO). 

We fine-tune Llama 3.2 as the base policy. This keeps the policy in the same model family as the generator used to produce candidate answer in phase~1, so that fine-tuning operates in the same distribution with respect to preference data. While for the judges, are deliberately drawn from other families to avoid self-preference bias since LLM evaluators systematically favour their own generations \citep{panickssery_llm_2024}.

\paragraph{Phase~3} 

We evaluate the baseline Llama-3.2 models against all four fine tuning methods (SFT, DPO, Dr. DPO, and GRPO), each trained on the full set of strong preference pairs. For DPO, we additionally train one model per corruption stratgey, isolating the effect of each corruption strategy on the learned preference. 

All models are evaluated on a held-out test of 100 questions, sampled randomly and stratified with a fixed random seed from previously unseen data, which are questions that not appear in any training data at all. For each model, we score the 100 generated answers with the judge ensamble and additionally report citation F1.

In conclusion, We evaluate and improve LLM performance on EU financial regulatory
question answering (QA) in three stages: (i) a \emph{pointwise
LLM-as-a-Judge} protocol that scores candidate answers along four
rubric dimensions; (ii) \emph{preference-based fine-tuning}, in which
judge scores are converted into preference pairs and further augmented
with targeted corruptions; and (iii) an \emph{evaluation strategy}
that combines the same judge ensemble with a rule-based Citation~F1
metric to compensate for the judges' known weakness on citation
quality.

To align the model toward citation faithfulness, we introduce an off-policy synthetic preference generation pipeline. Rather than relying on smaller student model's generations, we contruct preference pairs from ground-truth regulatory answers and high quality output from larger teacher models ( Apertus- 70B and Llama-3.3-70B). The targeted corruption taxonomy ensures that the student model learns from highly challenging distractors. 

\subsection{Pointwise Prompting with LLMs}
\label{sec:methods:pointwise}

We adopt a pointwise \emph{LLM-as-a-Judge} protocol rather than a
pairwise one. Pointwise scoring is known to be more reliable than
pairwise comparison for measuring scalar constructs and avoids the
position bias documented in pairwise setups
\citep{licht_measuring_2025,zeng_evaluating_2024}; the trade-off is
that pointwise judges are more vulnerable to adversarial responses
\citep{raina_is_2024}, which we exploit deliberately in
Section~\ref{sec:methods:finetuning}.

\paragraph{Rubric.}
Each candidate answer $a$ to a regulatory question $q$ (drawn from the
EBA and ESMA Q\&A corpora, with the official answer $a^\star$ as
reference) is scored on a 1--5 Likert scale along four dimensions:
\textbf{Accuracy}, \textbf{Completeness}, \textbf{Topic Coherence},
and \textbf{Citation Quality}. Each dimension is anchored by explicit
level descriptors (see Appendix~\ref{app:prompt-judge}); the overall score
is the arithmetic mean of the four.

\paragraph{Judge ensemble.}
To reduce single-model bias, every $(q, a, a^\star)$ triple is scored
independently by three open-weight judges:
\emph{Gemma-4-31B-it}, \emph{Qwen3.5-27B}, and
\emph{GLM-4.7-Flash}. Judges are prompted with the rubric, the
question, the reference answer, and the candidate answer, and are
required to return a JSON object with one sentence of reasoning per
dimension followed by an integer score. To avoid truncation
inside reasoning blocks, we set \texttt{max\_new\_tokens}~$=2048$ and
disable extended thinking mode where supported. The final pointwise
score for each dimension is the mean across the three judges.

\paragraph{Preference derivation.}
For every question, each of the three judges scores both candidate
answers (from \emph{Apertus-70B-Instruct} and
\emph{Llama-3.3-70B-Instruct}) independently. For every
(question, judge) pair we compute an \emph{overall} score as the
unweighted mean of the four rubric dimensions (Accuracy,
Completeness, Topic Coherence, Citation Quality), and record which
answerer that judge prefers. A pair is retained as a preference
example only if two conditions hold:
\begin{enumerate}[label=(\roman*), leftmargin=*, itemsep=1pt, topsep=2pt]
    \item \textbf{Unanimous agreement} : all three judges prefer
    the same answer (majority count $= 3$);
    \item \textbf{Strong preference} : at least one judge exhibits
    a score gap of $\geq 1.5$ between the two answers.
\end{enumerate}
The retained answer is labelled \emph{chosen} and the other
\emph{rejected}; all remaining pairs are discarded to reduce label
noise from ambiguous or judge-specific preferences.

Out of $N_{\text{total}} = 2{,}134$ questions, $858$ (\SI{40.2}{\percent})
yield unanimous judge agreement, of which $359$ additionally satisfy
the strong-preference criterion. We use these $359$ pairs as the
DPO training set and hold out an untouched set of $100$ pairs for
evaluation.

\begin{table}[h]
\centering
\caption{Preference pair construction. For each question, judges score both
answerer models on an overall score (mean of the four dimensions). A pair is
retained only if all three judges agree on the preferred answer and at least
one judge shows a strong preference (score gap $\ge 1.5$).}
\label{tab:pair_construction}
\begin{tabular}{lrr}
\toprule
\textbf{Criterion} & \textbf{Questions} & \textbf{\%} \\
\midrule
Total questions                          & 2{,}134 & 100.0 \\
Unanimous (all available judges agree)   & 913 & 42.8 \\
Unanimous (all 3 judges, majority $=3$)  & 858 & 40.2 \\
Unanimous \emph{and} strong preference   & 359 & 16.8 \\
\bottomrule
\end{tabular}
\end{table}

\begin{table}[h]
\centering
\caption{Composition of DPO-ready pairs ($n = 359$), by regulator and
preferred model.}
\label{tab:dpo_ready}
\begin{tabular}{lccr}
\toprule
& \multicolumn{2}{c}{\textbf{Preferred model}} & \\
\cmidrule(lr){2-3}
\textbf{Regulator} & Llama-3.3-70B & Apertus-70B & \textbf{Total} \\
\midrule
EBA   & 232 & 64 & 296 \\
ESMA  & 54  & 9  & 63 \\
\midrule
\textbf{Total} & 286 & 73 & 359 \\
\bottomrule
\end{tabular}
\end{table}

\subsection{Fine-tuning}
\label{sec:methods:finetuning}

\paragraph{Base model.}
We fine-tune \textsc{LLama 3.2-3B} as the student model. The 3B scale is
small enough to iterate over the full corruption grid on a single
GH200 GPU while remaining strong enough that gains reflect the
\emph{method}, not the base model.

\paragraph{Shared setup.} All four methods (SFT, DPO, Dr.DPO, and GRPO) use LoRA adapters (rank $r = 16$,
$\alpha = 32$) on all attention and MLP projections with 4-bit base weights,
\texttt{learning\_rate} $= 5\times10^{-6}$, $\beta = 0.1$, effective batch size
16, and 3 epochs. Checkpoints are selected on validation overall score.

\paragraph{Corruption strategies on DPO.}
All four methods (SFT, DPO, Dr. DPO, and GRPO) are trained on the full set of corrupted preference pairs. To isilate the contribution of each corruption type, we additionally train one DPO model per strategy, with the same objective and all hyperparameters fixed and varying only on which corruption types used on training data. This yields \textbf{four} additional models alongside the four main ones, for \textbf{eight} in total.

\subsection{Evaluation Strategy}
\label{sec:methods:evaluation}

We evaluate every fine-tuned checkpoint on a held-out test split of
$n=100$ EU regulatory questions, disjoint from the training and
validation splits. For each question we generate one answer from the
baseline (Llama3.2-3B) and one from the DPO checkpoint, then report
three families of metrics.

\paragraph{Judge scores.}
The same pointwise judge ensemble described in
Section~\ref{sec:methods:pointwise} scores each answer along the
four rubric dimensions. We report the mean across judges for each
dimension and the overall score.

\paragraph{Pairwise win rate.}
For each question, we compare the mean overall scores of the
baseline and DPO answers and record whether DPO is preferred,
baseline is preferred, or the two tie (equal to two decimal places).

\paragraph{Rule-based citation metric.}
Alongside the judge ensamble, we report a deterministic, rule based citation metric that is exactly reproducible and directly checkable against the source law. We extract the set of references cited in a generated answers and compare it against the gold reference set, reporting precision, recall and F1 over all exact article matches.

\subsubsection{Evaluation Metrics: Citation F1}
\label{sec:methods:citation_f1}

Citation~F1 treats each answer as a set of regulatory references and
compares it to the reference set extracted from the official answer
$a^\star$. References are extracted with a deterministic regular-
expression parser that recognises the instrument name
(e.g., \emph{CRR}, \emph{MiFIR}, \emph{DORA},
Regulation \emph{(EU) 2019/876}) and, where present, an article,
paragraph, sub-paragraph and point identifier.

We report Citation~F1 at two granularities:
\begin{itemize}
    \item \textbf{Instrument-level Citation~F1}: the citation is
    considered correct if the referenced regulation matches
    $a^\star$, regardless of article. This captures whether the
    model cites the right \emph{law}.
    \item \textbf{Article-level Citation~F1}: the citation is
    considered correct only if both the instrument and the article
    identifier match. This captures whether the model cites the
    right \emph{provision}.
\end{itemize}
For an answer $a$ with predicted citation set $C(a)$ and gold
citation set $C(a^\star)$, precision, recall and F1 are computed in
the standard way:
\begin{equation}
P = \frac{|C(a) \cap C(a^\star)|}{|C(a)|},
\quad
R = \frac{|C(a) \cap C(a^\star)|}{|C(a^\star)|},
\quad
F_1 = \frac{2 P R}{P + R}.
\end{equation}
Compared with judge-based Citation Quality, this metric is
deterministic, cheap to compute, and directly discriminates
\emph{fabricated} citations from \emph{correct} ones.
\section{Results}
\label{sec:results}

\subsection{Phase 1: Candidate Answer Quality and Judge Reliability}


\paragraph{Pointwise judge results.}

Both answerer models (Apertus 70B and Llama 3.3 70B) show the same pattern across all of the three judges. Topic coherence is consistenly high with score ranging from 4.12 until 4.57, while accuracy is much lower (2.13 to 2.40). The patterns holds on both EBA and ESMA subsets. In other words, the answerer models stay on topic and write clearly but with some incomplete substance compared to the ground truth. 

\paragraph{Apertus-70B underperforms Llama-3.3-70B despite matched scale.} At the same 70B parameter count, Apertus consistently underperforms Llama-3.3 on every rubric dimension, with per-metric gaps of 0.2--0.5 points. LLama-3.3 has been trained more extensively to follow instructions and produce well-formed answers, which makes it better at following the task. Thus,  this suggests that the gap comes from post-training rather than capacity. Which means, two models on the same size can differ substantially in output quality, especially in specialized EU financial regulatory QA.



\begin{figure}[H]    \includegraphics[width=1\linewidth]{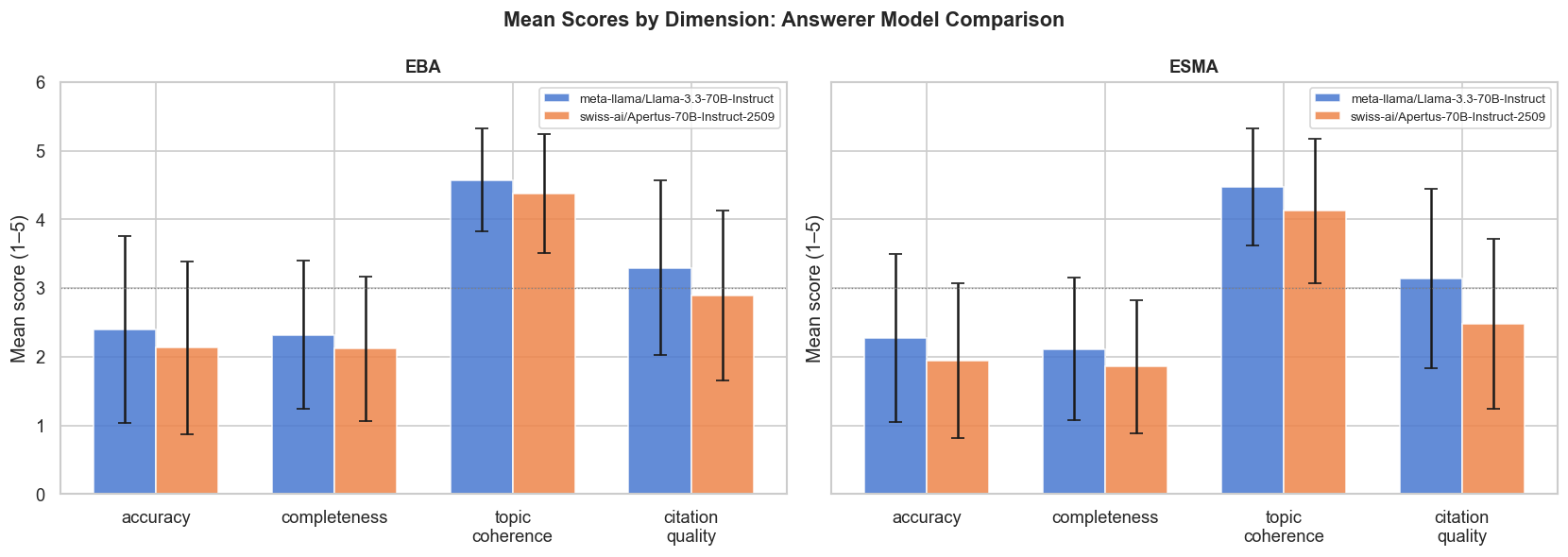}
    \caption{Mean judge scores (1-5) by answerer model and regulator, avaraged across three judges}
    \label{fig:placeholder}
\end{figure}
\begin{table}
\centering
\caption{Mean judge scores (1--5) by answerer model and regulator,
averaged across the three judges.}
\label{tab:answerer_scores}
\begin{tabular}{llcccc}
\toprule
\textbf{Regulator} & \textbf{Model} & \textbf{Accuracy} & \textbf{Completeness} & \textbf{Coherence} & \textbf{Citation} \\
\midrule
EBA  & Llama-3.3-70B & \textbf{2.40} & \textbf{2.32} & \textbf{4.57} & \textbf{3.29} \\
     & Apertus-70B   & 2.13 & 2.12 & 4.38 & 2.89 \\
\midrule
ESMA & Llama-3.3-70B & \textbf{2.27} & \textbf{2.11} & \textbf{4.47} & \textbf{3.14} \\
     & Apertus-70B   & 1.94 & 1.86 & 4.12 & 2.48 \\
\bottomrule
\end{tabular}
\end{table}
\paragraph{Inter-judge agreement between all three judges} For our pipeline, what matters is that all three judges agree on the same pattern. By evaluating the pairwise inter-judge agreement results using Krippendorff\'s $\alpha$, it falls between 0.4-0.7 across all dimensions, indicating moderate to substantial agreement between all three judges except for Qwen-GLM pairs on citation quality. The agreement is strong in accuracy (0.68--0.78) and completeness (0.7--0.74) between all three pairs of judges and moderate in topic coherence (0.40--0.56). Citation quality is one of the exception, while Qwen-Gemma and Gemma-GLM pairs are still in a workable range, Qwen-GLM agreement collapse to 0.014, which means that the two judges apply different standard for what counts as a correct citation, which is precisely why we complement with the deterministic rule-based citation metric.

\subsection{Phase 2: Preference Pair Construction}

\begin{table}[h!]
\centering
\caption{Pairwise inter-judge agreement (Krippendorff's $\alpha$).}
\label{tab:judge_agreement}
\begin{tabular}{lccc}
\toprule
& \multicolumn{3}{c}{\textbf{Judge pair}} \\
\cmidrule(lr){2-4}
\textbf{Dimension} & Qwen--Gemma & Qwen--GLM & Gemma--GLM \\
\midrule
Accuracy         & 0.781 & 0.682 & 0.738 \\
Completeness     & 0.737 & 0.695 & 0.703 \\
Topic coherence  & 0.558 & 0.455 & 0.399 \\
Citation quality & 0.345 & 0.014 & 0.413 \\
\bottomrule
\end{tabular}
\end{table}

\paragraph{Preference derivation.}
From the pointwise judge scores, we derive preference pairs between the two answerer models per the criteria of §\ref{sec:methods:pointwise}. As expected, the largest chosen--rejected gaps occur on Accuracy and Completeness, the same dimensions on
which the models struggle most. While ties are more frequent on Topic Coherence, where both models score highly. This confirms that the preference signal is concentrated on the axes we most want the
fine-tuned model to improve, which is accuracy and completeness. 

\begin{table}[H]
\centering
\caption{Mean absolute score gap $|s_\text{chosen} - s_\text{rejected}|$ per
dimension, over DPO-ready pairs. Accuracy drives the preference signal,
topic coherence contributes least.}
\label{tab:gap_by_dim}
\begin{tabular}{lc}
\toprule
\textbf{Dimension} & \textbf{Mean gap (1--5)} \\
\midrule
Accuracy         & 1.998 \\
Completeness     & 1.461 \\
Citation quality & 1.218 \\
Topic coherence  & 0.928 \\
\bottomrule
\end{tabular}
\end{table}

\paragraph{What separates the chosen from the rejected response.} Table \ref{tab:gap_by_dim} reports the average score difference between $|s_\text{chosen} - s_\text{rejected}|$ on each rubric dimension. Accuracy shows the widest gap (1.998 points), followed by completeness (1.461), while topic coherence shows the narrowest (0.928). In other words, the preference pairs are mostly separated by whether an answer is correct and complete, not by how well it stays on topic, both responses in a pair usually address the question. 

\paragraph{Separation before and after corruption} The unanimous judge filter alone already yields well-separated pairs, the mean gap between $|s_\text{chosen} - s_\text{rejected}|$ for accuracy is close to two points out of five. The Phase~2 corruptions widen this separation further. Swapping a law or an article makes the rejected responses verifiably wrong. In addition, injecting fabricated citations or texts also makes the rejected responses even worse. Thus, filtering and corruption together yield 359 pairs with a sting and verifiable quality gap, which we use as the training set in Phase~2.

\subsection{Phase 3: Fine-Tuning and Evaluation}

\begin{table}
  \centering
  \begin{tabular}{lcc}
    \toprule
    Method & Overall & $\Delta$ vs.\ baseline \\
    \midrule
    SFT              & 2.281 & $+0.322$ \\
    DPO              & 2.085 & $+0.113$ \\
    Dr.\ DPO         & 2.074 & $+0.101$ \\
    GRPO             & 2.056 & $+0.088$ \\
    \bottomrule
  \end{tabular}
  \caption{Overall judge score on the held-out test set, with improvement over
  the untuned Llama~3.2 baseline. All methods trained on the full set of
  strong preference pairs.}
  \label{tab:method-comparison}
\end{table}

\begin{table}
  \centering
  \setlength{\tabcolsep}{5pt}
  \begin{tabular}{lccccc}
    \toprule
    Method & Accuracy & Completeness & Coherence & Cite quality & Overall \\
    \midrule
    SFT      & 1.590 \up{0.023} & 1.740 \up{0.313} & 3.620 \up{0.530} & 2.173 \up{0.420} & 2.281 \up{0.322} \\
    DPO      & 1.589 \up{0.026} & 1.529 \up{0.092} & 3.391 \up{0.271} & 1.832 \up{0.065} & 2.085 \up{0.113} \\
    Dr.\ DPO & 1.609 \up{0.042} & 1.522 \up{0.082} & 3.368 \up{0.245} & 1.796 \up{0.036} & 2.074 \up{0.101} \\
    GRPO     & 1.607 \up{0.045} & 1.493 \up{0.059} & 3.357 \up{0.243} & 1.767 \up{0.007} & 2.056 \up{0.088} \\
    \bottomrule
  \end{tabular}
  \caption{Per-dimension judge scores after fine-tuning, with the change from
  the untuned Llama~3.2 baseline shown in green. All methods trained on the
  full set of strong preference pairs.}
  \label{tab:per-dimension}
\end{table}

\paragraph{Fine-tuning setup.} We fine-tune in two configurations. First, we train all four methods (SFT, DPO, Dr.~DPO, GRPO) on the combined set of corrupted preference pairs. Second, we train DPO alone on each corruption type separately, as an ablation to identify which corruption contributes most to the observed gains.

\paragraph{Overall judge scores.} The untuned Llama~3.2 baseline scores 1.959 overall. We report 
$\Delta$ ( Table \ref{tab:method-comparison})as the difference between each fine-tuned model and this baseline. All four methods improve on it, with SFT gaining the most (~+0.322) and GRPO the least (~+0.088). The per-dimension (Table \ref{tab:per-dimension}) breakdown shows where SFT's advantage comes from: almost entirely from citation quality (+0.420, while on accuracy it remains below Dr.~DPO and GRPO. Accuracy is the dimension we most want to improve in this setting, which GRPO get the higgest accuracy score improvement.

\paragraph{Judges reward length and citation density.} Based on table \ref{tab:results-full}, SFT produces markedly longer answers than the other methods (299.7 words against 200--237) and cites far more sources (5.11 per answer against 1.59--1.93). The judges score these answers highly on citation quality, apparently reading a dense reference list as evidence of grounding. While it got a high score in judge citation quality, the deterministic metric disagrees : SFT's citation F1 is 0.181, the lowest of all models. Thus, The additional citations produces on SFT answers are largely hallucinated, see \ref{app:sftvsgrpo} for detailed examples.

\paragraph{Win rate.} Alongside the mean judge score, we report also the win rate in Table \ref{tab:results-full}, which is the fraction of test questions on which a judge prefers the fine-tuned model's answer to the baseline's. SFT wins on 68\% of questions, while DPO, Dr.~DPO, and GRPO sit at 0.49--0.50. Although SFT gets the highest win rate, the rule-based metric gives the opposite picture : SFT has the lowest citation F1 (0.181), and DPO and GRPO have the highest among the fine-tuned models (0.229).

\paragraph{Implications for regulatory QA.} This divergence (Table \ref{tab:results-full}) is the central finding in our evaluation. A judge-only score would rank SFT first, but the rule-based metric F1 Score shows that it is the least reliable model for regulatory use. DPO and GRPO offer the better trade off with more concise answer and highest citation among all the fine-tuned models.




\begin{table}
  \centering
  \setlength{\tabcolsep}{5pt}
  \begin{tabular}{lccccc}
    \toprule
    Method & Overall Judge Score & Win rate & Citation F1 & Citations/answer & Length (words) \\
    \midrule
    SFT      & 2.281 & \textcolor{deltagreen}{0.68} & \textcolor{deltared}{0.181} & \textcolor{deltared}{5.11} & \textcolor{deltared}{299.7} \up{38.1} \\
    DPO      & 2.085 & 0.49 & \textcolor{deltagreen}{0.229} & 1.59 & 236.9 \\
    Dr.\ DPO & 2.074 & 0.50 & 0.223 & 1.93 & 222.5 \\
    GRPO     & 2.056 & 0.49 & \textcolor{deltagreen}{0.229} & 1.79 & 200.2 \\
    \bottomrule
  \end{tabular}
  \caption{Judge scores, rule-based citation results, and answer length on the
  held-out test set. SFT's higher judge score coincides with longer answers, a
  sharp rise in citation count, and a drop in citation F1 below the baseline.
  All fine-tuned models fall below the baseline citation F1 of 0.245.}
  \label{tab:results-full}
\end{table} 

\paragraph{Per-corruption ablation.} Based on Table \ref{tab:corruption-ablation}, In the second configuration we train DPO on each corruption type separately, holding all other settings fixed, to identify which corruption contributes most to the observed gains. Against the untuned Llama~3.2 baseline of 1.959, all four models show a slight overall improvement (+0.016 to +0.053). Law swap performs best on both overall judge score (2.027) and citation F1 (0.251), while win rates are close across all four (0.43--0.45).

\begin{table}
  \centering
  \begin{tabular}{lcccc}
    \toprule
    Corruption & Overall & $\Delta$ & Win rate & Citation F1 \\
    \midrule
    Law swap              & \textcolor{deltagreen}{2.027} & $+0.053$ & 0.43 & \textcolor{deltagreen}{0.251} \\
    Hallucinated text     & 2.001 & $+0.027$ & 0.44 & 0.225 \\
    Hallucinated citation & 1.999 & $+0.027$ & 0.45 & 0.230 \\
    Article swap          & 1.989 & $+0.016$ & 0.44 & 0.231 \\
    \bottomrule
  \end{tabular}
  \caption{Per-corruption DPO ablation on the held-out test set. Each model is
  trained with DPO on a single corruption strategy, with all other settings
  held fixed. Row order is by overall judge score.}
  \label{tab:corruption-ablation}
\end{table} 
\FloatBarrier              
\section{Conclusion}

We introduce a novel benchmark for financial regulatory question answering, a domain where correctness and precise citation are non-negotiable. Candidate answers are generated by two 70B open-weight models (Llama-3.3-70B and Apertus-70B) and scored pointwise by three judges drawn from different model families (Qwen3.5, GLM-4.7, and Gemma-4). Retaining only unanimously judged pairs and augmenting the rejected side with four corruption strategies : law swap, article swap, hallucinated citation, and hallucinated text, yields preference pairs with a wide and verifiable quality gap, suitable for preference training in this domain.

Three findings as follow. \textbf{First}, the answerer models struggle on accuracy and completeness while scoring consistently high on topic coherence. Thus, staying on topic is easy, but being correct is harder. \textbf{Second}, SFT achieves the highest overall judge score but the lowest citation F1, because its answers are markedly longer and carry nearly three times as many citations, most of them unsupported. The judges read citation density as grounding, but the deterministic metric shows otherwise. Thus, DPO and GRPO offer the better trade-off with concise answers and the highest citation F1 among fine-tuned models. \textbf{Third}, in the per-corruption ablation, law swap yields the highest overall score and citation F1 of the four strategies. 

We release the benchmark and evaluation stack, including the rule-based citation metric, to support citation-aware evaluation of models proposed for compliance use.

\clearpage

\bibliographystyle{plainnat}
\bibliography{references}

\clearpage

\appendix

\section{Datasets}
\label{app:dataset-appendix}

\subsection{Dataset Generation Result Statistics}
\label{app:dataset-stat}

\begin{table}[h]
\centering
\caption{Pairwise inter-judge agreement. $\kappa_\text{lin}$ and
$\kappa_\text{quad}$ are linear- and quadratic-weighted Cohen's $\kappa$;
$\alpha$ is Krippendorff's alpha.}
\label{tab:judge_agreement_full}
\begin{tabular}{llrrrr}
\toprule
\textbf{Dimension} & \textbf{Judge pair} & $n$ & $\kappa_\text{lin}$ & $\kappa_\text{quad}$ & $\alpha$ \\
\midrule
\multirow{3}{*}{Accuracy}
  & Qwen--Gemma & 4{,}276 & 0.648 & 0.785 & 0.781 \\
  & Qwen--GLM   & 4{,}183 & 0.548 & 0.692 & 0.682 \\
  & Gemma--GLM  & 4{,}183 & 0.596 & 0.738 & 0.738 \\
\midrule
\multirow{3}{*}{Completeness}
  & Qwen--Gemma & 4{,}276 & 0.631 & 0.740 & 0.737 \\
  & Qwen--GLM   & 4{,}183 & 0.547 & 0.696 & 0.695 \\
  & Gemma--GLM  & 4{,}183 & 0.552 & 0.704 & 0.703 \\
\midrule
\multirow{3}{*}{Topic coherence}
  & Qwen--Gemma & 4{,}276 & 0.426 & 0.559 & 0.558 \\
  & Qwen--GLM   & 4{,}183 & 0.341 & 0.458 & 0.455 \\
  & Gemma--GLM  & 4{,}183 & 0.305 & 0.405 & 0.399 \\
\midrule
\multirow{3}{*}{Citation quality}
  & Qwen--Gemma & 4{,}276 & 0.277 & 0.432 & 0.345 \\
  & Qwen--GLM   & 4{,}183 & 0.130 & 0.239 & \textbf{0.014} \\
  & Gemma--GLM  & 4{,}183 & 0.327 & 0.435 & 0.413 \\
\bottomrule
\end{tabular}
\end{table}

\begin{figure}[H]
    \centering
    \includegraphics[width=1\linewidth]{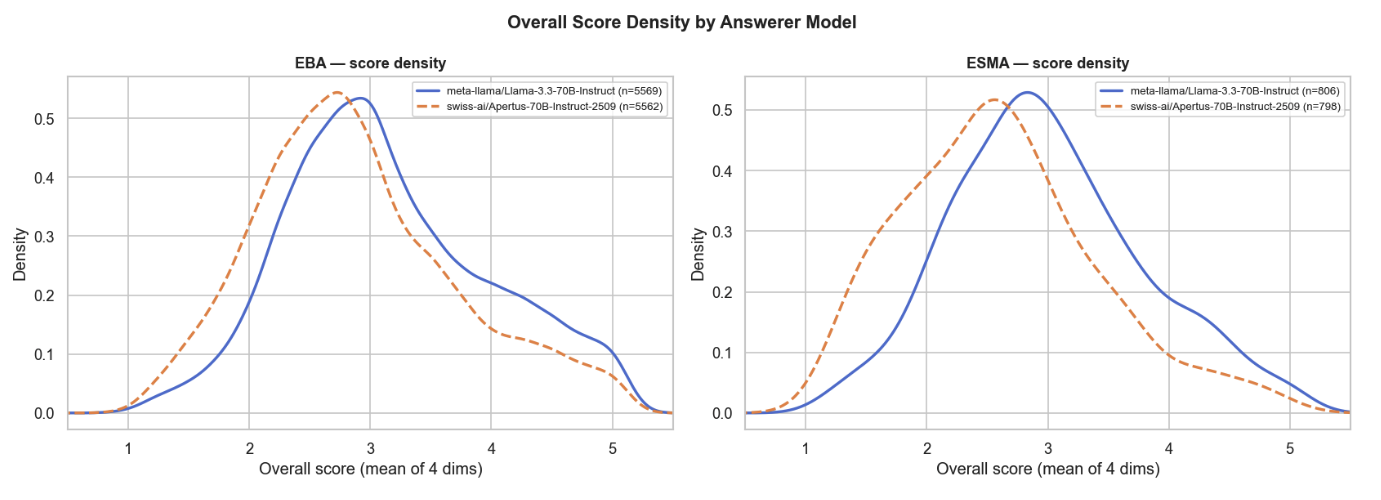}
    \caption{Score Distribution across answerer model }
    \label{fig:placeholder}
\end{figure}

\newpage

\section{Prompts}
\label{app:prompts}
 
This appendix contains the full text of the prompts used in our pipeline.
Placeholders enclosed in curly braces (e.g.\ \texttt{\{legal\_act\}}) are
filled at inference time with the corresponding fields of the evaluation
instance.
 
\subsection{Answer Generation Prompt}
\label{app:prompt-answerer}
 
The answer generation prompt elicits an EBA/ESMA-style response to a
regulatory question conditioned on the structured context of the
instance (legal act, topic, subject matter, and background).
 
\begin{lstlisting}[style=promptstyle]
You are an expert in EU financial regulation, with deep knowledge of EBA and ESMA guidelines, technical standards, and related directives and regulations.
 
You will receive a regulatory question along with structured context:
- LEGAL ACT: the specific instrument the question concerns
- TOPIC: the regulatory area
- SUBJECT MATTER: the specific provision, field, or requirement
- BACKGROUND: the context or inconsistency motivating the question
 
## Requirements
 
1. **ACCURACY**: Base your answer on the actual content of the legal act identified above and any directly relevant secondary instruments (RTS, ITS, guidelines). Address the specific subject matter and background; do not answer a more general version of the question.
 
2. **CITATIONS**: Every substantive claim must be supported by a specific citation in the form [Source, Article/Paragraph/Field], e.g., [Regulation (EU) 2022/2554, Art. 28(3)] or [EBA/ITS/2023/01, field B_05.01.0020]. The primary citation should point to the legal act named in the context. General references such as "DORA" or "the ITS" are not sufficient. Use square brackets consistently.
 
3. **ABSTENTION**: If you cannot identify a specific provision for a claim, write "[no specific provision identified]" immediately after the claim, rather than fabricating a citation. Partial answers with honest gaps are preferred over complete answers with fabricated references.
 
4. **FORMAT**: 100-400 words of prose, matching the style of official EBA/ESMA Q&A responses. Inline citations after each substantive claim. Do not pad.
 
5. **CONCISENESS**: State each point once. Do not restate the question, do not write a closing "In summary" / "In conclusion" paragraph that repeats what you already said, and do not hedge with filler like "it is worth noting" or "however, it should be noted". If you do not know something, say so plainly in one sentence rather than reasoning around it at length.
 
## Context
 
LEGAL ACT: {legal_act}
TOPIC: {topic}
SUBJECT MATTER: {subject_matter}
BACKGROUND: {background}
 
## Question
 
{question}
 
Answer :
\end{lstlisting}
 
\subsection{Judge Prompt}
\label{app:prompt-judge}
 
The judge prompt scores a candidate answer against the official EBA/ESMA
answer on four dimensions (\emph{accuracy}, \emph{completeness},
\emph{topic coherence}, and \emph{citation quality}) using a five-point
rubric, and returns a structured JSON object with one reasoning sentence
and one integer score per dimension.
 
\begin{lstlisting}[style=promptstyle]
You are an expert judge evaluating answers to EU financial regulation questions from EBA and ESMA sources. You will score a candidate answer against the official answer on four dimensions.
 
## Question
{question}
 
## Topic
{topic}
 
## Subject Matter
{subject_matter}
 
## Official Answer (Ground Truth)
{ground_truth}
 
## Candidate Answer
{candidate}
 
## Scoring Dimensions
 
**Accuracy** (factual correctness relative to the official answer):
- 5: All factual claims match the official answer.
- 4: Minor inaccuracies that do not change the substantive conclusion.
- 3: Partially correct; one substantive claim is wrong or unsupported.
- 2: Multiple substantive errors; conclusion is partly incorrect.
- 1: Conclusion contradicts the official answer or is fabricated.
 
**Completeness** (coverage of key points in the official answer):
- 5: Covers all key points present in the official answer.
- 4: Covers all key points but omits a minor detail.
- 3: Covers the main point but misses one secondary point.
- 2: Misses multiple key points; partial coverage.
- 1: Misses the main point entirely.
 
**Topic Coherence** (alignment with the specified topic and subject matter):
- 5: Fully addresses the specified topic and subject matter without drifting into adjacent regulatory areas.
- 4: Stays on topic but includes minor tangential content, OR addresses a broader scope that still fully covers the topic.
- 3: Partially on topic; some content addresses a different but related regulatory area.
- 2: Primarily addresses an adjacent topic; only partially relevant to the specified subject matter.
- 1: Off-topic or addresses a different regulatory area entirely.
 
**Citation Quality** (specificity and correctness of legal references):
- 5: All citations are specific (article/paragraph/field level) and correctly identify the supporting provision.
- 4: Citations are specific and mostly correct; one minor citation issue.
- 3: Citations are present but partially generic (e.g., "Article 5" without specifying the regulation), or one citation is incorrect.
- 2: Citations are mostly generic ("the ITS", "DORA") or several are incorrect.
- 1: Citations are missing or fabricated.
 
## Instructions
 
1. **Reason before scoring.** Think step-by-step about each dimension before assigning the score. Your reasoning should determine the score, not the reverse.
 
2. **Catch plausible-sounding hallucinations.** If the candidate contains specific factual claims (numbers, dates, article references, requirements, definitions) that are not supported by the official answer, treat those as accuracy violations even if the claims sound plausible.
 
3. **Score dimensions independently.** If two dimensions seem to conflict, score each on its own merits.
 
4. **Use the full 1-5 range.** Do not default to 3 when uncertain. A shorter candidate that still covers the key point is not a completeness penalty.
 
4b. **Score must match your own reasoning.** If your reasoning sentence for a dimension names a specific flaw (a missing point, a wrong citation, an unsupported claim, drift off-topic), the score for that dimension cannot be 5. Reserve 5 only when your reasoning states the candidate fully and correctly meets the criterion with no caveat. Do not default to 5 out of leniency.
 
5. **Citation specificity matters.** A correct citation to "Article 5 of Regulation (EU) 2019/2033" is stronger than "Article 5" alone, even when both are technically correct.
 
6. **Be concise.** Do your step-by-step thinking silently. Each reasoning field must be ONE short sentence (max ~25 words) stating the conclusion, not a transcript of your deliberation. Do not repeat the question, the candidate text, or the official answer back. Output the JSON object immediately after you have decided the scores -- no preamble, no text before or after the JSON.
 
7. **No visible thinking.** Do not output a `<think>` block, chain-of-thought, or any reasoning outside the JSON's "reasoning" fields. Your entire response must be the JSON object and nothing else.
 
Respond in this exact JSON format only:
{
  "reasoning": {
    "accuracy": "<one sentence>",
    "completeness": "<one sentence>",
    "topic_coherence": "<one sentence>",
    "citation_quality": "<one sentence>"
  },
  "accuracy": <1-5>,
  "completeness": <1-5>,
  "topic_coherence": <1-5>,
  "citation_quality": <1-5>
}
\end{lstlisting}

\section{Examples of SFT vs GRPO}
\label{app:sftvsgrpo}

\begin{figure}[H]
\small
\begin{qualbox}{Question}
Does Article~97(1)(b) PSD2 apply to pay-by-invoice when the payer's funds are
covered by a credit line extended by a payment service provider?
\end{qualbox}

\begin{qualbox}{Ground truth}
Article~97(1)(b) PSD2 requires strong customer authentication (SCA) for
payer-initiated payment transactions. Article~4(5) defines a payment
transaction irrespective of the underlying obligation between payer and payee,
and Article~4(3) read with point~4 of Annex~I identifies the execution of
payment transactions covered by a credit line as a payment service. SCA
therefore applies.
\end{qualbox}

\begin{qualbox}{{SFT (415 words, 7 citations)}}
\textit{[Abridged.]} States that Art.~97(1)(b) requires SCA where funds are
\emph{not} covered by a credit line, then argues the provision ``does not
explicitly address'' the credit-line case. Infers an answer from
Recitals~20--21 and Art.~3(4), suggests consulting the EBA or ESMA for
clarification, and cites EBA/GL/2020/01 \S7.8 twice as offering only general
guidance. Concludes tentatively that SCA ``should be applied''.
\end{qualbox}

\begin{qualbox}{{GRPO (24 words, 1 citation)}}
Yes. Article~97(1)(b) PSD2 applies to pay-by-invoice transactions where the
payer's funds are covered by a credit line extended by a payment service
provider.
\end{qualbox}

\caption{Qualitative comparison on a held-out PSD2 question. The SFT model
produces a longer, heavily cited answer that inverts the rule, treats the
provision as silent on the case at hand, and defers to further consultation.
The GRPO model answers directly and correctly. Judges score the SFT answer
higher on citation quality.}
\label{fig:qualitative-example}
\end{figure}

\newpage

\section{Corruption Strategies: Worked Example}
\label{app:corruption_examples}

To make the corruption strategies of
Section~\ref{sec:methods:finetuning} concrete, we illustrate each of
the five perturbations on a single EBA Q\&A item concerning the use
of non-EUR currency equivalents for EUR thresholds under the RTS. In
every case, the \emph{chosen} answer is the reference regulator
answer; only the \emph{rejected} side is corrupted. Corrupted spans
are highlighted in \textcolor{red!70!black}{red}; original spans
retained from the reference answer are shown in
\textcolor{gray!70!black}{gray}.

\begin{tcolorbox}[
    colback=gray!5,
    colframe=gray!40,
    title=\textbf{Source item (EBA Q\&A)},
    fonttitle=\small\bfseries,
    boxrule=0.4pt, arc=1pt, left=4pt, right=4pt, top=3pt, bottom=3pt
]
\small
\textbf{Question.} May payment service providers (PSPs) and card
schemes set rounded and easily understandable non-EUR currency
equivalents for the EUR thresholds set out in the RTS?

\smallskip
\textbf{Reference answer (excerpt).} ``According to
\textit{Regulation~(EU)~2022/2554, Art.~28(3)}, the thresholds for
exemptions in the RTS are set in EUR\ldots''
\end{tcolorbox}

\vspace{4pt}

\paragraph{(1) \texttt{law\_swap}.}
The article number is preserved, but the cited instrument is
replaced with a topically adjacent but incorrect regulation or
directive. This targets failure modes in which the model recalls a
plausible article number under the wrong legal instrument.

\begin{quote}
\small
``According to
\textcolor{red!70!black}{\textit{Directive 2011/61/EU (AIFMD)}},
\textcolor{gray!70!black}{Art.~28(3), the thresholds\ldots}''
\end{quote}

\paragraph{(2) \texttt{article\_swap}.}
The cited instrument is preserved, but the article identifier is
replaced with a different article of the \emph{same} instrument.
This targets the more subtle failure mode of citing the correct law
but the wrong provision.

\begin{quote}
\small
``According to
\textcolor{gray!70!black}{\textit{Regulation~(EU)~2022/2554},
Art.~\textcolor{red!70!black}{126(3)}, the thresholds\ldots}''
\end{quote}

\paragraph{(3) \texttt{hallucinate\_citation}.}
A fabricated citation --- one whose regulation number and article do
not exist --- is appended to an otherwise fluent, non-committal
answer. This targets the fabricated-citation pattern observed
frequently in the baseline outputs (Section~\ref{sec:analysis}).

\begin{quote}
\small
\textcolor{gray!70!black}{``\ldots[no specific provision
identified].} This conclusion is further reinforced by
\textcolor{red!70!black}{\textit{Regulation~(EU) No~9138/2039,
Article~731}}, which explicitly mandates additional disclosure
requirements not otherwise referenced here.''
\end{quote}

\paragraph{(4) \texttt{hallucinate\_text}.}
No fabricated citation is introduced; instead, a plausibly worded
but factually unsupported regulatory claim is inserted. This
isolates \emph{content} hallucination from \emph{citation}
hallucination.

\begin{quote}
\small
\textcolor{gray!70!black}{``\ldots[no specific provision
identified].} It should also be noted that
\textcolor{red!70!black}{the relevant competent authority may, at
its sole discretion, exempt branches with fewer than 50 employees
from this obligation}.''
\end{quote}

\paragraph{(5) \texttt{combination}.}
A random subset of the above four strategies is applied jointly to
the same answer, producing rejected examples in which multiple
failure modes co-occur.

\begin{quote}
\small
``According to
\textcolor{red!70!black}{\textit{Regulation~(EU) No~1093/2010},
Art.~90(3)}\ldots This conclusion is further reinforced by
\textcolor{red!70!black}{\textit{Regulation~(EU) No~9829/2070,
Article~841}}, which explicitly mandates additional disclosure
requirements not otherwise referenced here.
\textcolor{red!70!black}{This obligation does not apply during the
first three years following the entity's initial
authorisation}.''
\end{quote}

\vspace{4pt}
\noindent
Table~\ref{tab:corruption_summary} summarises which failure mode
each strategy is designed to target.

\begin{table}[H]
\centering
\small
\setlength{\tabcolsep}{5pt}
\caption{Summary of corruption strategies and the failure mode each
one targets.}
\label{tab:corruption_summary}
\begin{tabular}{lll}
\toprule
\textbf{Strategy} & \textbf{Perturbation} & \textbf{Targeted failure mode} \\
\midrule
\texttt{law\_swap}            & Instrument $\to$ different law, article kept & Wrong law, plausible article \\
\texttt{article\_swap}        & Article $\to$ different article, law kept    & Right law, wrong provision \\
\texttt{hallucinate\_citation}& Fabricated regulation/article inserted        & Non-existent citations \\
\texttt{hallucinate\_text}    & Fabricated regulatory claim inserted         & Content hallucination without citation \\
\texttt{combination}          & Random subset of the above four              & Co-occurring failure modes \\
\bottomrule
\end{tabular}
\end{table}

\end{document}